\documentclass[aps,twocolumn,epsfig,groupedaddress,showpacs,floatfix]{revtex4}
                                                                                
\usepackage{color}
                                                                                
\usepackage{graphicx}
\usepackage{epsfig}
                                                                                
\usepackage{amssymb}
\usepackage{amsmath}

\def\menorsim{\smash{\mathop{<}\limits_{\raise3pt\hbox{$\sim$}}}}
\def\maiorsim{\smash{\mathop{>}\limits_{\raise3pt\hbox{$\sim$}}}}

\begin{document}


                                                                                

\title{A model for net-baryon rapidity distribution}



\author{J. Alvarez-Mu\~niz}
\affiliation{IGFAE and Dep. Fisica Particulas, Univ. Santiago de Compostela, 
15782 Santiago de Compostela, Spain}
\author{R. Concei\c{c}\~ao}
\author{J. Dias de Deus}
\author{M.C. Esp\'{\i}rito Santo}
\author{J. G. Milhano}
\author{M. Pimenta}
\email[Corresponding author]{, pimenta@lip.pt}
\affiliation{CENTRA, LIP and IST, Av. Rovisco Pais, 1049-001 Lisboa, Portugal}


\date{\today}

\begin{abstract}
In nuclear collisions, a sizable fraction of the available energy 
is carried away by baryons. As the baryon number is conserved, the net-baryon $B-\bar{B}$ 
retains information on the energy-momentum carried by the incoming nuclei.
A simple and consistent model for net-baryon production in high energy proton-proton 
and nucleus-nucleus collisions is presented.
The basic ingredients of the model are valence string formation based on standard PDFs 
with QCD evolution and string fragmentation via the Schwinger mechanism.
The results of the model are presented and compared with data at different
centre-of-mass energies and centralities, as well as with existing models.
These results show that a good 
description of the main features of net-baryon data is possible in the framework of a 
simplistic model, with the advantage of making the fundamental production mechanisms manifest.
\end{abstract}




\maketitle

\section{\label{sec:level1}Introduction}
\label{sec:introd}

In hadron-hadron, hadron-nucleus and nucleus-nucleus interactions a sizable fraction of the available energy 
in a collision is carried away by baryons~\cite{netbar,vanhove,feynman}. 
As the baryon number is conserved, the measured net-baryon, $B-\bar{B}$, 
keeps track of the energy-momentum carried by the incoming particles. 
An important question to be asked is: how does the fraction of energy carried by the net-baryon evolve as a function 
of the centre-of-mass collisional energy per nucleon, $\sqrt{s}$~?
This question is important because, on one hand, a decrease of the fraction of energy going into the net-baryon implies more 
energy available to create the deconfined quark-gluon state of matter, and on the other hand, such a decrease may reduce the 
possibility of producing fast particles in very high energy cosmic ray experiments.
For more than 30 years, since the ISR at CERN, particle production studies have been limited to mid rapidity. Fortunately, 
with RHIC, large rapidity data became available, and, hopefully, the same will happen with the LHC.
In fact, if one does not measure the physics at high rapidity the most elementary physical constraint, namely energy conservation, 
cannot be applied~\cite{GM-JDD,Back}.

In most of the existing Monte Carlo models~\cite{qgsjet1,qgsjet2,epos,sibyll}, 
the physics of net-baryon production is very much obscured by the complexity of extensive 
and detailed codes. Often, the basic production mechanisms do not appear in a transparent way. 
In this paper we present a simple and consistent model for net-baryon production in high energy proton-proton (p-p)
and nucleus-nucleus (A-A) collisions.
Preliminary versions of this model have been presented in~\cite{ournetB}. 
As it happens in most of the existing codes, we shall work in the framework of the Dual Parton Model (DPM)~\cite{capella} 
with string formation (valence strings, in the present case) based on standard Parton Distribution Functions (PDFs) 
with QCD evolution and string fragmentation via the Schwinger mechanism.
The basic ingredients of the model are:
\begin{itemize}
\item Formation of extended color fields or strings, making use of PDFs for valence quarks;
\item Evolution with momentum scale $Q^2$, as a consequence of the QCD evolution of the PDFs;
\item Fragmentation of each string with formation of a fast baryon (net-baryon) and other particles.
\end{itemize}
With this simple model we show that a significant part of the physics of net-baryon production can
be understood in terms of valence strings and $Q^2$ evolution, while sea strings play a less relevant role.

When going from p-p to A-A collisions,
it must be noted that the net-baryon results depend directly on the number of participants in the collision
and thus on the collision centrality.
The relation between the number of participants and the impact parameter of the collision can be established, 
for instance, in the context of the Glauber model~\cite{glauber}. 
For A-A collisions, centrality thus plays an important role in our model.

This paper is organized as follows. In section~\ref{sec:current}, the available experimental data and 
the predictions of current models 
are briefly reviewed and compared. In section~\ref{sec:themodel}, our net-baryon model is described. The results of 
the model are presented and compared with data and with existing models in section~\ref{sec:results}. 
A summary and brief conclusions are presented in section~\ref{sec:conclusions}.

\section{Review of current model predictions and experimental data}
\label{sec:current}

The data presently available on net-proton or net-baryon rapidity is relatively scarce. 
The existing experimental results used in this study are briefly reviewed here.

The most recent data are from the BRAHMS collaboration at RHIC~\cite{rhic-brahms,rhic-brahms2}, which published 
net-proton and net-baryon rapidity distribution for central Au-Au collisions at $\sqrt{s}=200$~GeV and $\sqrt{s}=62.4$~GeV. 
The 5\% most central collisions were used at $\sqrt{s}=200$~GeV, while at $\sqrt{s}=62.4$~GeV the 10\% most central 
collisions were selected.
The procedure to obtain the net-baryon distributions from the measured net-proton ones (properly accounting for
strangeness and for neutrons) is detailed in the references. 
In the BRAHMS net-proton distributions, weak decay corrections 
(to remove the contribution from protons/antiprotons produced not at vertex but from strange baryon decays) 
were not applied.

At the SPS (Pb-Pb collisions at $\sqrt{s} \simeq 17$~GeV), the NA49 collaboration published net-proton rapidity distributions
with weak decay corrections included 
for central collisions (the 5\% most central collisions were selected)~\cite{sps}. 
Net-proton results for different centrality bins 
have been presented by this experiment in~\cite{sps2}.
In this work data have been split into five centrality bins: 0-5\%, 5-14\%, 14-23\%, 23-31\%, 31-48\% and 48-100\%.

At lower centre-of-mass energies, net-proton rapidity distributions have been obtained at AGS (Au-Au at 
$\sqrt{s} \simeq 5$~GeV)~\cite{ags} for the most central collisions. Centrality cuts of 
5\%, 4\% and 3\% were used by the E917, E877 and E802 experiments, respectively. 
Weak decay corrections are negligible at AGS energies.

\begin{figure}[htb]
\begin{center}
\fontsize{9pt}{9pt}
\epsfig{figure=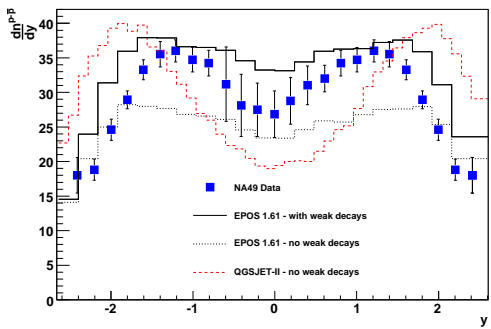, angle=0, width=0.4\textwidth}
\caption{The net-proton rapidity distributions from EPOS 1.61 (full and dotted lines) and QGSJET-II.03 (dashed line) for 
Pb-Pb collisions at $\sqrt{s} \simeq 17$~GeV are shown and compared with NA49 data (points)
with weak decay corrections~\cite{sps}.
For EPOS, the results obtained including (full line) and excluding (dotted line) the
strangeness contribution are shown. For QGSJET, the `no-decays' curve is shown.}
\label{netbar_mod}
\end{center}
\end{figure}

The experimental results are usually presented in terms of net-proton and net-baryon rapidity $y$ distributions
and in terms of rapidity loss, 
defined as
\begin{equation}
\left< \delta y  \right> = y_p - \left< y \right>\, ,
\end{equation}
where $y_p$ is the beam rapidity and $\left< y \right>$ is the mean net-baryon rapidity after the collision, 
given by
\begin{center}
\begin{equation}
\left< y \right> = \frac{2}{N_{part}} \int _{0}^{y_p} y \frac{dn^{B-\bar{B}}(y)}{dy} dy\, .
\label{mean_yb}
\end{equation}
\end{center}
Here, $N_{part}$ is the number of participants in the collision and $N_{B-\bar{B}}$ is the net-baryon number.

Let us now turn to net-baryon production as implemented in the existing Monte Carlo models.
QGSJET-II~\cite{qgsjet2} and EPOS~\cite{epos} are amongst the presently most widely used hadronic 
models in high energy and cosmic ray physics.
To our knowledge, there is no systematic study comparing the predictions of these two models on 
net-baryon production between themselves or with experimental data. 

In figure~\ref{netbar_mod}, net-proton rapidity distributions obtained with QGSJET-II.03 and EPOS 1.61
for Pb-Pb collisions at $\sqrt{s} \simeq 17$~GeV are shown and compared with experimental data. 
According to~\cite{sps} weak decay corrections have been applied to the data.
Following~\cite{Pb-centrality}, the impact parameter range (0 to 3.1 fm) corresponding to the centrality
cut of 5\% applied to data was selected in QGSJET and EPOS. 
While in EPOS an impact parameter cut is performed directly, in QGSJET this is implemented
through a multiplicity cut~\cite{qgsjet2}.
For EPOS, the results obtained leaving all particles free to decay and contribute 
to the net-baryon are shown together with those obtained switching off all weak decays.
The `all-decays' EPOS curve reproduces the trend seen in data, with an excess that could
be due to the presence of net-baryon from weak decays. However, the strangeness
effect compensation in the `no-decays' curve seems to be too strong.
For QGSJET-II, the `no-decays' curve is shown and was found to be very similar to what is 
obtained from the model by default. It should be noted that in QGSJET the parton shower is fully developed
before hadronisation is performed, and  the connection to the experimental picture is less direct.
One should recall that QGSJET-II is not expected to perform very well at such low energies.

\section{The model}
\label{sec:themodel}

\begin{figure}[htb]
\begin{center}
\fontsize{9pt}{9pt}
\epsfig{figure=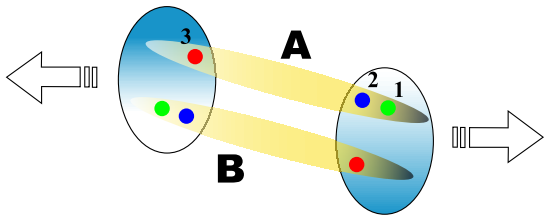, angle=0, width=0.4\textwidth}
\caption{Schematic representation of a proton-proton collision, with the formation of two valence strings.}
\label{collision}
\end{center}
\end{figure}

In the spirit of~\cite{vanhove,feynman,capella,bass},
the basic assumption of our model
is that net-baryon production in proton-proton collisions is strongly correlated with 
the formation and fragmentation of two color singlet strings, each one with two valence quarks from one of the protons, 
and one valence quark from the other proton. This is schematically shown in figure~\ref{collision}. 

Referring to the figure,
let $x_1$, $x_2$ and $x_3$ be the fractions of momentum carried by the valence quarks forming string A. 
Quarks 1 and 2 are from the proton with positive momentum in the proton-proton reference frame, 
and quark 3 is from the other proton. No transverse momentum is considered within this model.
By choice, $x_1 > x_2 > x_3$, with $x_3<0$. 
The energy and momentum of each string are obtained adding directly the energy and momentum 
carried by each of the valence quarks. For string A:
\begin{center}
\begin{eqnarray}
E_{string} &=& (x_1 + x_2 + (-x_3))\frac{\sqrt{s}}{2}\, ,\\
P_{string} &=& (x_1 + x_2 - (-x_3))\frac{\sqrt{s}}{2}\, ,\\
M_{string} &=& \sqrt{(x_1 + x_2)\ (-x_3)\ s}\, ,
\end{eqnarray}
\end{center}
where the quark momentum fractions $x_1$, $x_2$ and $x_3$ are determined from the valence 
quark PDFs at an effective momentum scale $Q^2$.
For each $\sqrt{s}$, an effective $Q^2$ derived from a fit to experimental data will be chosen
(see section~\ref{sec:results}).
In this work the CTEQ6M parton distribution functions~\cite{cteq6} were used.

\begin{figure}[htb]
\fontsize{9pt}{9pt}
\begin{center}
\epsfig{figure=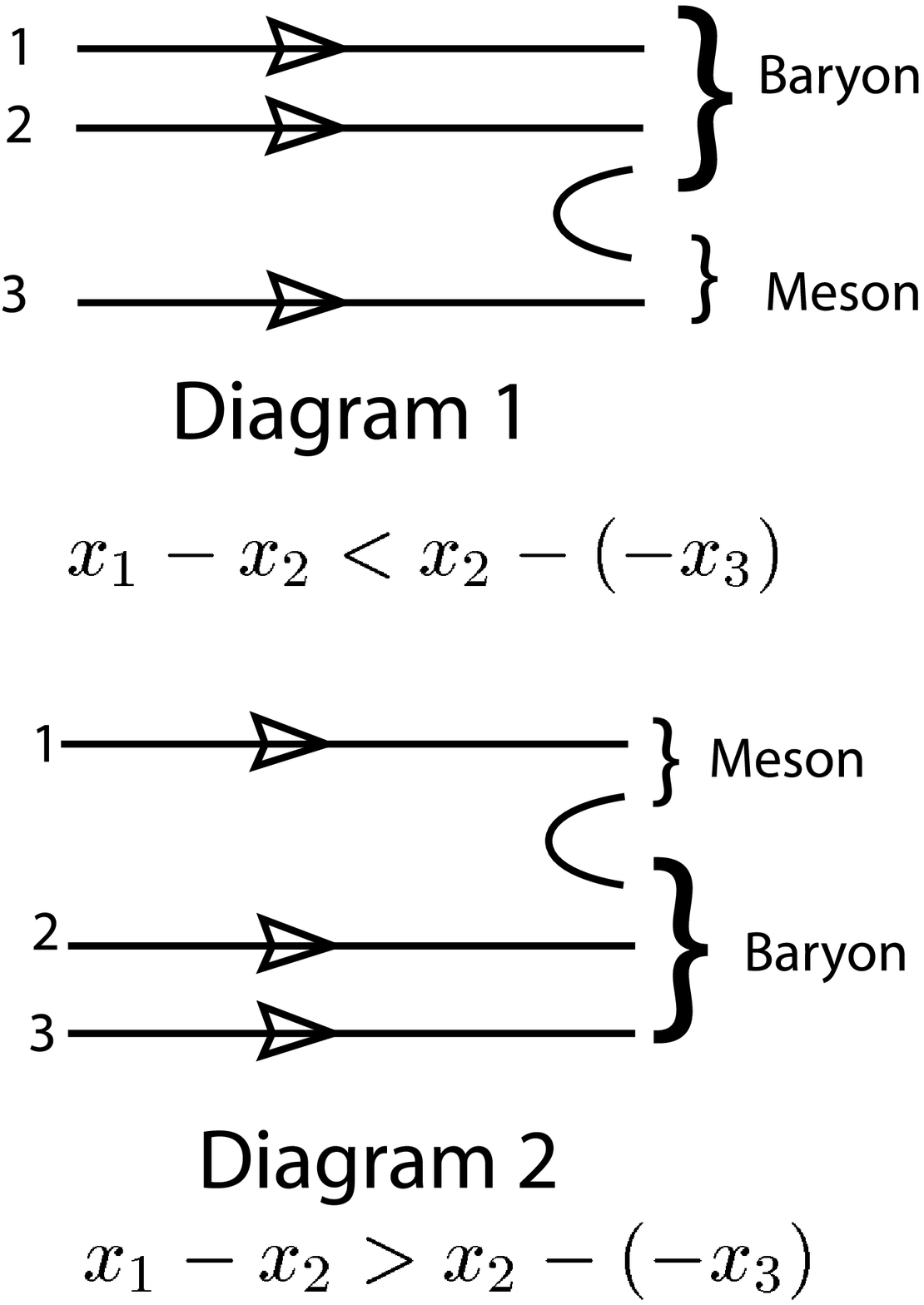, angle=0, width=0.3\textwidth}
\caption{The two main valence string fragmentation diagrams.
In the simple model assumed, the string is cut in two pieces and a $q \bar{q}$ pair is formed, from the vacuum,
between the two quarks with the largest momentum difference.
Diagram 1 corresponds to the case $x_1 - x_2 < x_2 - (-x_3)$, in which quarks 2 and 3 (belonging to different protons)
are chosen. In Diagram 2, $x_1 - x_2 > x_2 - (-x_3)$ and string fragmentation occurs between quarks 1 and 2
(belonging to the same proton).} 
\label{dpmff}
\end{center}
\end{figure}

The simplest possible model for fragmentation is assumed. Each string decays into a baryon and a meson in the following way:
the string is cut in two pieces and a $q \bar{q}$ pair is formed, from the vacuum, either between quarks 2 and 3 (belonging
to different protons) or between quarks 1 and 2 (belonging to the same proton, with positive momentum in the case of string A).
The quark pair with the largest momentum difference is chosen. 
The string piece that inherits two valence quarks originates the baryon, whereas the string piece that inherits one valence 
quark originates the meson. 
This mechanism corresponds to the diagrams represented in figure~\ref{dpmff}. 
Diagram 1 corresponds to the case $x_1 - x_2 < x_2 - (-x_3)$, in which quarks 2 and 3 are chosen.
In Diagram 2, $x_1 - x_2 > x_2 - (-x_3)$ and string fragmentation occurs between quarks 1 and 2.
The weights of the two diagrams are, 
in this model, given only by kinematics. 
For string A the first diagram will be more probable (especially for low $\sqrt{s}$). However, the weight of the second diagram 
can be as large as $40\%$ above LHC energies.

The $q \bar{q}$ pair formed from the vacuum was taken to be either a $u \bar{u}$ or a $d \bar{d}$,
and the full quark combinatorics was then performed in order to determine the nature of the possible 
outcoming baryon. Both fundamental and excited states were considered, taking spin-dependent weights (2j+1). 
The decays of the unstable baryons were then performed and the outcoming nucleons included in the 
net-baryon calculations.
The contribution from $s$ quarks was not considered.
The strangeness effect was estimated from the Schwinger model to be about 25\% to 35\%. 
The implications of this approximation are discussed below.

\begin{figure}[htb]
\begin{center}
\fontsize{9pt}{9pt}
\epsfig{figure=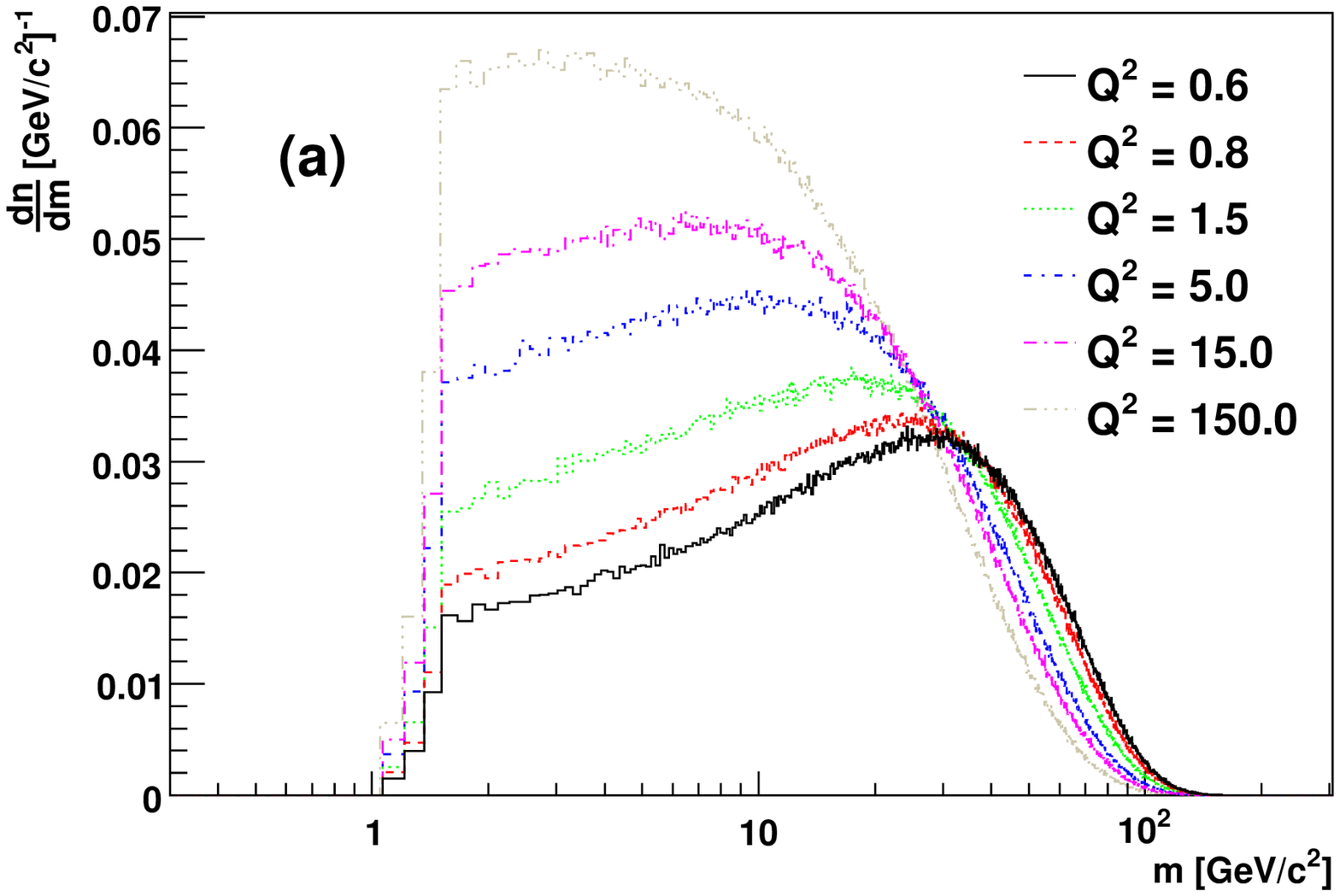, angle=0, width=0.4\textwidth}
\epsfig{figure=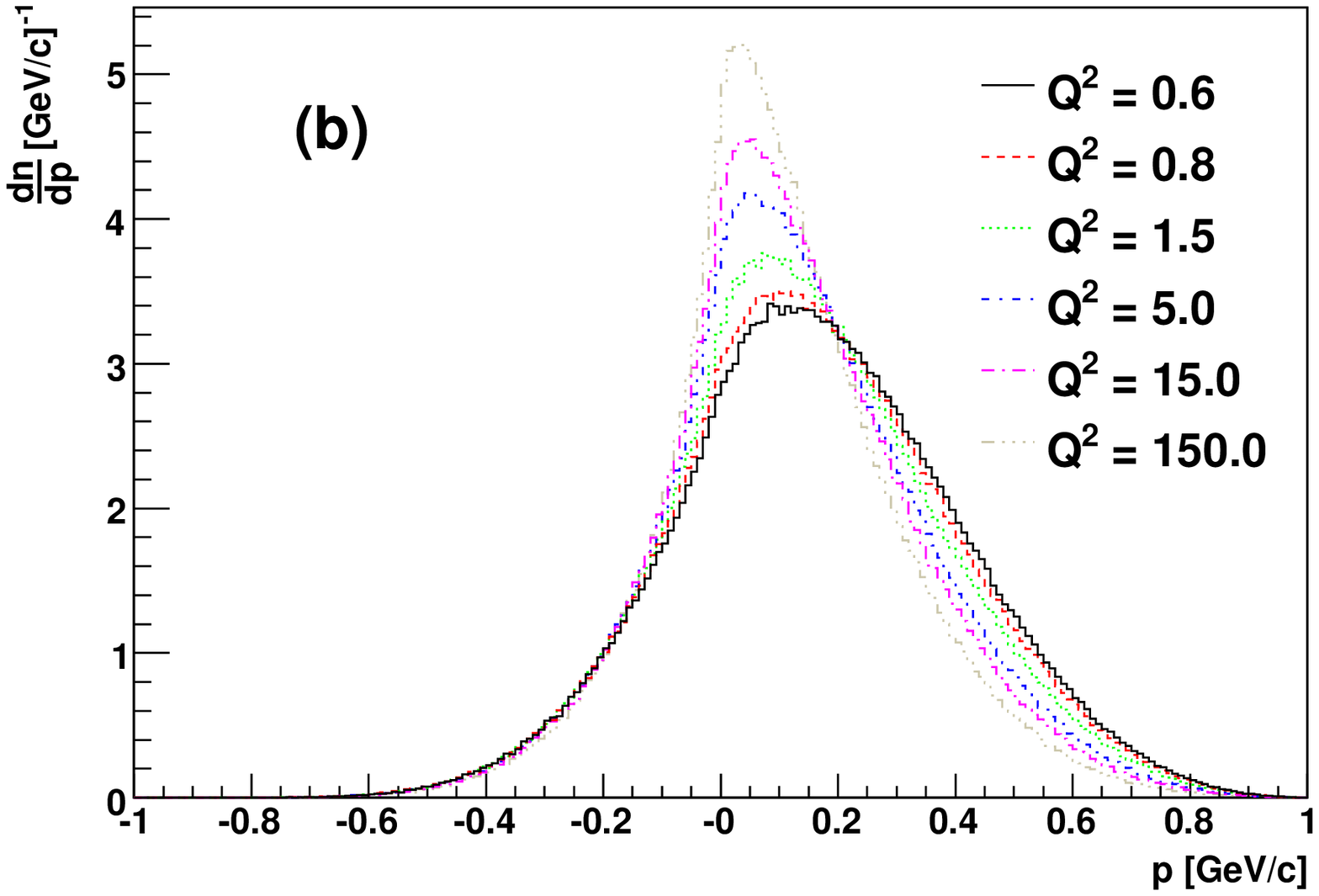, angle=0, width=0.4\textwidth}
\caption{Evolution with $Q^2$ (in (GeV/$c$)$^2$) of (a) the string mass and (b) the string momentum
distributions for proton-proton collisions $\sqrt{s}=200$~GeV.
The mass distribution applies to both valence strings, A and B (see figure~\ref{collision}). 
The momentum distribution is for the case of string A, and is symmetric for string B.}
\label{string200}
\end{center}
\end{figure}

\begin{figure}[htb]
\begin{center}
\fontsize{9pt}{9pt}
\epsfig{figure=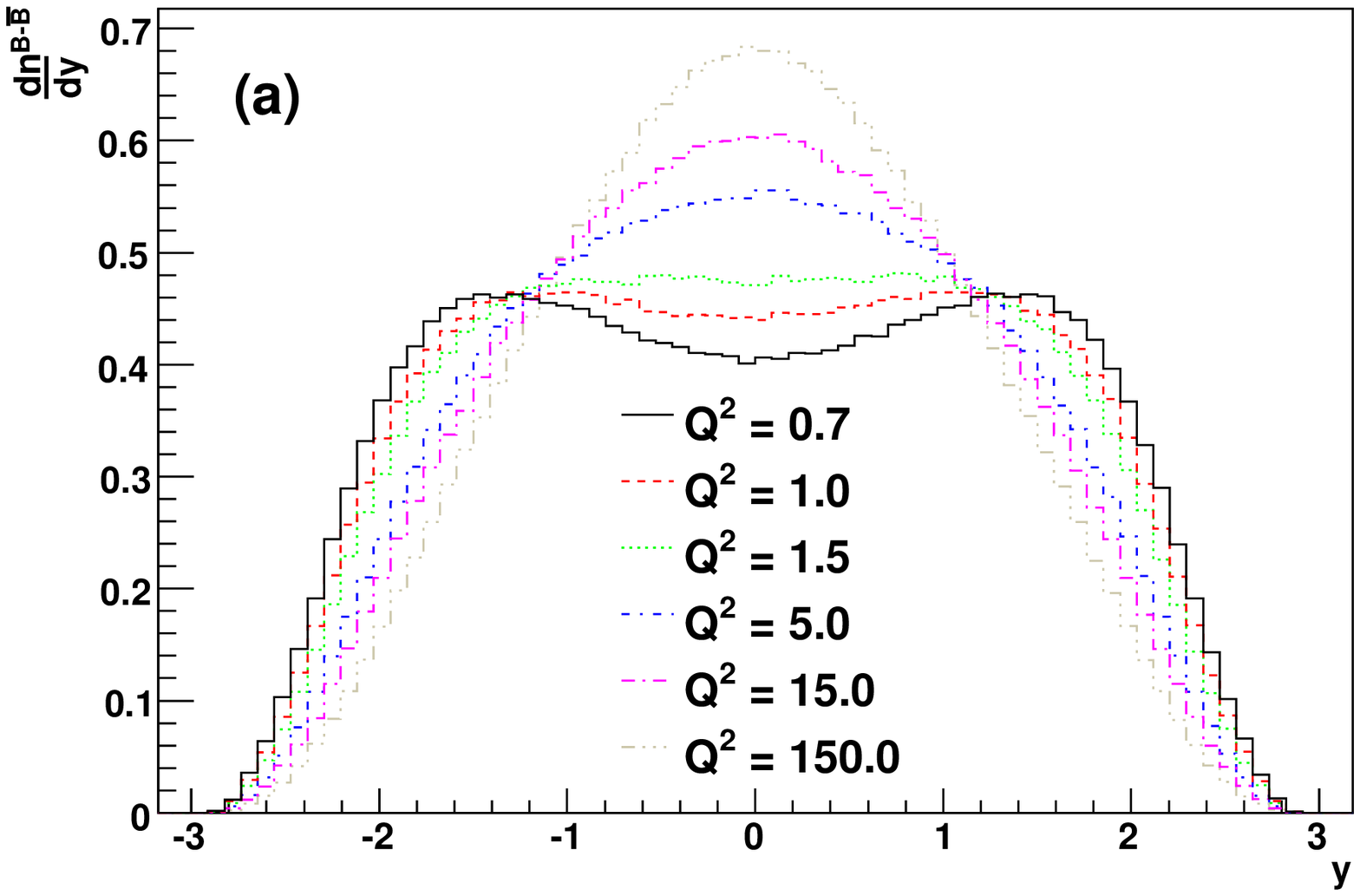, angle=0, width=0.4\textwidth}
\epsfig{figure=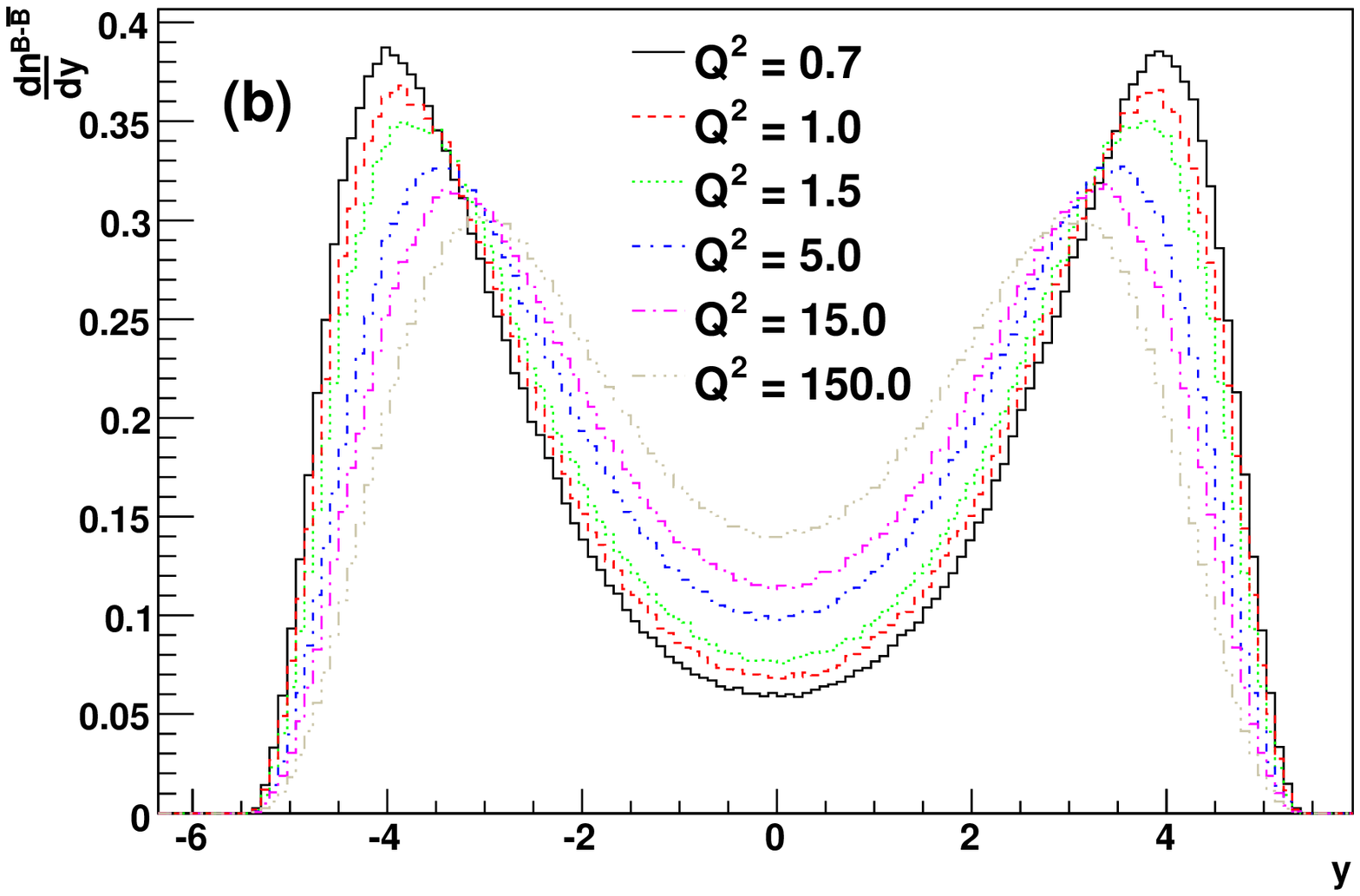, angle=0, width=0.4\textwidth}
\caption{Evolution of the net-baryon rapidity with $Q^2$ (in (GeV/$c$)$^2$) for proton-proton collisions 
at (a) $\sqrt{s}=17$~GeV and (b) $\sqrt{s}=200$~GeV.}
\label{netbar}
\end{center}
\end{figure}

The string mass and momentum distributions as a function of $Q^2$ are given in figure~\ref{string200}. 
The mass distribution applies to both valence strings, while the momentum distribution for the case of 
string B is simply symmetric.
The net-baryon rapidity distributions for different values of $Q^2$
are presented in figure~\ref{netbar}, for proton-proton collisions at two different centre-of-mass energies. 
This figure illustrates not only the evolution with $Q^2$ but also
the clear effect of kinematics, due to the increase of $\sqrt{s}$, 
on the main features of the distribution. It should be noted that, as we are dealing with
valence strings, when the $Q^2$ increases the mean rapidity decreases, simply due to PDF evolution.

It is worth stressing that the inclusion of diagram 2, in addition to diagram 1, with weights determined
simply by kinematics, reproduces some of the effects predicted in models with string 
junctions~\cite{junctions,shabelski} or popcorn~\cite{popcorn} mechanisms for the transport of baryon number 
from the beam rapidity into the central region $y \sim 0$. 
These effects can thus be achieved in a simple DPM model, based
on valence strings and a $Q^2$ parameterisation.

It is thus apparent that in this model the net-baryon rapidity $dn/dy^{B-\bar{B}}$ in proton-proton collisions is obtained 
from the rapidity of the two baryons produced in the fragmentation of the valence strings,
having the effective momentum scale $Q^2$ at each $\sqrt{s}$ as the only free parameter.

Let us now consider the case of A-A collisions. As noted above, the net-baryon results depend 
directly on the collision centrality. The number of participants will thus play an important role in 
the model. We shall assume that the net-baryon rapidity in A-A collisions at a 
given $\sqrt{s}$ can be obtained from the net-baryon rapidity in p-p collisions at the same $\sqrt{s}$
computed at an effective $Q^2_A$ value which depends
both on $\sqrt{s}$ and on A, with a normalisation factor which is
the number of participants per nucleus, $N_{part}/2$:

\begin{equation}
\label{eq:AA-netB}
\frac{dn}{dy} ^{B-\bar{B}} (Q^2_A) \Bigr \rvert_{A-A} \simeq \frac {N_{part}}{2} 
\cdot \frac{dn}{dy}^{B-\bar{B}} (Q^2_A) \Bigr \rvert _{p-p},
\end{equation}
where both rapidity distributions are evaluated at $Q^2_A$.
As stated above, the model does not include $s \bar{s}$ pairs.
We are thus assuming that strangeness does not considerably distort the distribution.

When comparing to net-proton results, we consider that net-proton is 
roughly 1/2 of $(B - \bar{B})$~\cite{rhic-brahms}. 
As strangeness effects are not included in the model, the obtained rapidity distribution
is well suited for comparing to data with weak decay corrections included. 
The strangeness effect is in this case simply a global factor $\varepsilon_s$,
estimated from the Schwinger model to be about 25\% to 35\%, which can be 
included in the normalisation factor: 
\begin{equation}
\label{eq:AA-netp}
\frac{dn}{dy} ^{p-\bar{p}} (Q^2_A) \Bigr \rvert_{A-A} \simeq 
\frac{N_{part}}{4} \cdot (1-\varepsilon_s) \cdot \frac{dn}{dy} ^{B-\bar{B}} (Q^2_A) \Bigr \rvert _{p-p}.
\end{equation}

Nuclear effects correction factors for the valence quark PDFs were estimated using EKS98~\cite{eks98} and 
nDS~\cite{nds} and found to be below 10-15\%. These corrections are taken into
account in the calculations.

To address the energy evolution of the model, a relation between the effective $Q^2$ and $\sqrt{s}$ needs to be established.
The effective $Q^2$ corresponds to the typical transverse size (area) 
of the parton (here, the valence quark). It is reasonable to assume, as in
Regge phenomenology~\cite{regge}, that the average number of partons in a nucleon
increases as a power of the centre of mass energy $\sqrt{s}$. Thus, 
$R_h^2  Q^2 \sim \sqrt{s}/\sqrt{s_0}$, 
where $R_h$ is the nucleon radius which we take as fixed. It then follows that
$Q^2$ should grow, allowing for deviations from this na\"{i}ve expectation, according to
\begin{equation}
 Q^2 = Q_0^2 \left( \frac{s}{s_0} \right)^{\lambda_v} \left[ GeV^2 \right].
\label{eq:q-lambda}
\end{equation}
Further, the effective A-A $Q^2_A$ should be analogously related to $\sqrt{s}$, but account for an extra dependence on the number of participants. Since the number of partons involved in an A-A collision grows proportionally to the number of nucleons from each nucleus involved in the collision (i.e. $\propto  
{N_{part}}/{2}$) and the transverse size (area) of the nucleus grows proportionally to $ ({N_{part}}/{2})^{2/3}$, we can expect the typical inverse size of a parton $Q^2_A$ to grow proportionally to $ ({N_{part}}/{2})^{1/3}$. Also, we can expect this simple geometrical estimate to be modified by nuclear effects and thus write for the A-A effective $Q_A^2$, as a function of the p-p effective $Q^2$:
\begin{equation}
\label{eq:q-alpha}
Q_A^2  = \left( \frac{N_{part}}{2} \right) ^\alpha Q^2.
\end{equation}
The value of $\alpha$ will be estimated using the data at different
centralities available for $\sqrt{s} \simeq 17$~GeV.

Combining equations~(\ref{eq:q-lambda}) and~(\ref{eq:q-alpha}), we can thus write, for A-A collisions:
\begin{equation}
 Q^2_A =  Q_0^2 \left( \frac{N_{part}}{2} \right)^\alpha \left( \frac{s}{s_0} \right)^{\lambda_v} \left[ GeV^2 \right],  
\label{eq:q-all}
\end{equation}
where the exponent $\lambda_v$ will be estimated performing a fit to the available data points.
Below, the free parameters of the model will be fixed 
by adjusting the results of the model to the experimental data
at each $\sqrt{s}$ and centrality. 


\section{Results}
\label{sec:results}

In this section, the available data on net-baryon and net-proton rapidity reviewed
in section~\ref{sec:current} will be used to fix the free parameters of the model.
The ability of the model to describe these data in a consistent way will
be evaluated and discussed.
While the data at different centralities allows to study the dependence on
the number of participants and to address the relation between
p-p and A-A, data at different centre-of-mass energies allow to establish and check
the evolution with $\sqrt{s}$ of the model predictions.

\begin{figure}[htb]
\begin{center}
\fontsize{9pt}{9pt}
\epsfig{figure=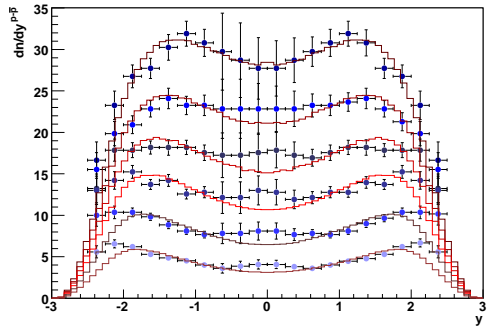, angle=0, width=0.4\textwidth}
\caption{The results of the present model for net-proton rapidity 
are compared to NA49 data in different centrality ranges 
(from top to bottom: 0-5\%, 5-14\%, 14-23\%, 23-31\%, 31-48\% and 48-100\%.)~\cite{sps2}.}
\label{na49-dndy}
\end{center}
\end{figure}

The experimental results are presented in terms of 
net-baryon rapidity in the case of BRAHMS
and in terms of net-proton rapidity in the case of NA49
and of the AGS experiments.
The model calculations are performed using equation~(\ref{eq:AA-netB}) for the net-baryon case
and~(\ref{eq:AA-netp}) for the net-proton one.  

The NA49 net-proton rapidity distributions for different 
centrality ranges at $\sqrt{s} \simeq 17$~GeV presented in~\cite{sps2}
were used in the first step of this analysis.
Our model was fitted to these data points in each centrality bin,
taking as free parameters the effective $Q^2_A$ and a normalisation factor $n$. 
The data points and the results of the model are shown in 
figure~\ref{na49-dndy}. A satisfactory agreement is achieved for all the centrality ranges.
The apparent difference between data and the model in the mid-rapidity region, 
could be due to the fact that the experimental error bars do not include the error 
associated to weak corrections.
On the other hand, difractive effects are not included in the model.
This accounts for the poorer description of the
high rapidity extremes in peripherial collisions, where such effects 
are expected to become more relevant.
It is worth noting that the effect seen in data displacing 
the maxima of the distribution to higher rapidity values as
the centrality of the collision decreases, is reproduced in the model
through a decrease of the fitted $Q^2_A$ value.

\begin{figure}[htb]
\begin{center}
\fontsize{9pt}{9pt}
\epsfig{figure=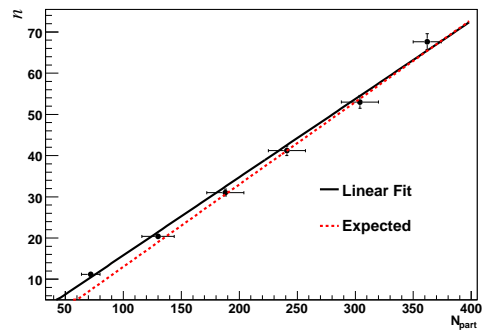, angle=0, width=0.4\textwidth}
\caption{
Results of the fit to the NA49 rapidity distributions at different centralities:
the fitted and expected value of the normalisation factor $n$ are plotted against 
the value of $N_{part}$ computed in~\cite{sps2} for each centrality bin.
}
\label{na49-norm}
\end{center}
\end{figure}

The normalisation factor $n$ obtained from the fit can be compared with what
would be expected from equation~(\ref{eq:AA-netp}) using the number of participants estimated 
by the experiment in~\cite{sps2} and assuming $\varepsilon_s=0.25$.
This comparison is shown in figure~\ref{na49-norm}, where the fitted and expected 
values of $n$ are plotted against the value of $N_{part}$ computed
in the reference for each centrality bin.
We see that a good agreement is found. The relation between $n$ and $N_{part}$
is thus well understood,
giving us confidence on the model and on the fitting procedure.

\begin{figure}[htb]
\begin{center}
\fontsize{9pt}{9pt}
\epsfig{figure=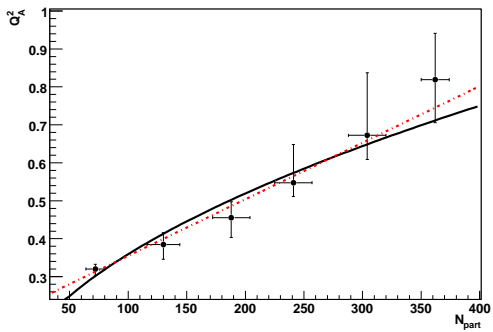, angle=0, width=0.4\textwidth}
\caption{
Results of the fit to the NA49 dn/dy distributions at different centralities:
the obtained $(Q^2_A,N_{part})$ points are shown and fitted with equation~(\ref{eq:q-alpha})
(full) and with a straight line (dashed).
}
\label{na49-npart}
\end{center}
\end{figure}

We thus have, for each centrality, a number of participants
and a value of $Q^2_A$.
These results can be used to estimate the exponent $\alpha$ on
equation~(\ref{eq:q-alpha}), which relates the effective $Q^2$ in p-p and A-A 
collisions. This was done by fitting equation~(\ref{eq:q-alpha}) to these
$(Q^2_A,N_{part})$ points, leaving $Q^2$ and $\alpha$ as free
parameters. The data points are shown in figure~\ref{na49-npart}, together
with the fit result, which corresponds to $\alpha=0.53 ^{+0.12}_{-0.13}$.
A linear fit to these points is also plotted, showing that the 
limited range covered in $N_ {part}$ does
not allow to strongly constrain the form of the function describing 
this dependence. The constant term added to $Q^2$ at $N_{part}$=2 
in the case of the linear fit is however hard to motivate. 
It should be noted that a simple interpolation between
the points is sufficient to relate $Q^2_A$ and $Q^2$
in the range of $N_{part}$ relevant for the interpretation of
the existing data.
The exact form of the function would however
become relevant when extrapolating to p-p collisions.
The uncertainty  on the relation  between $Q^2$ and $Q^2_A$ 
is thus at present too large to allow such an extrapolation.

\begin{figure}[htb]
\begin{center}
\fontsize{9pt}{9pt}
\epsfig{figure=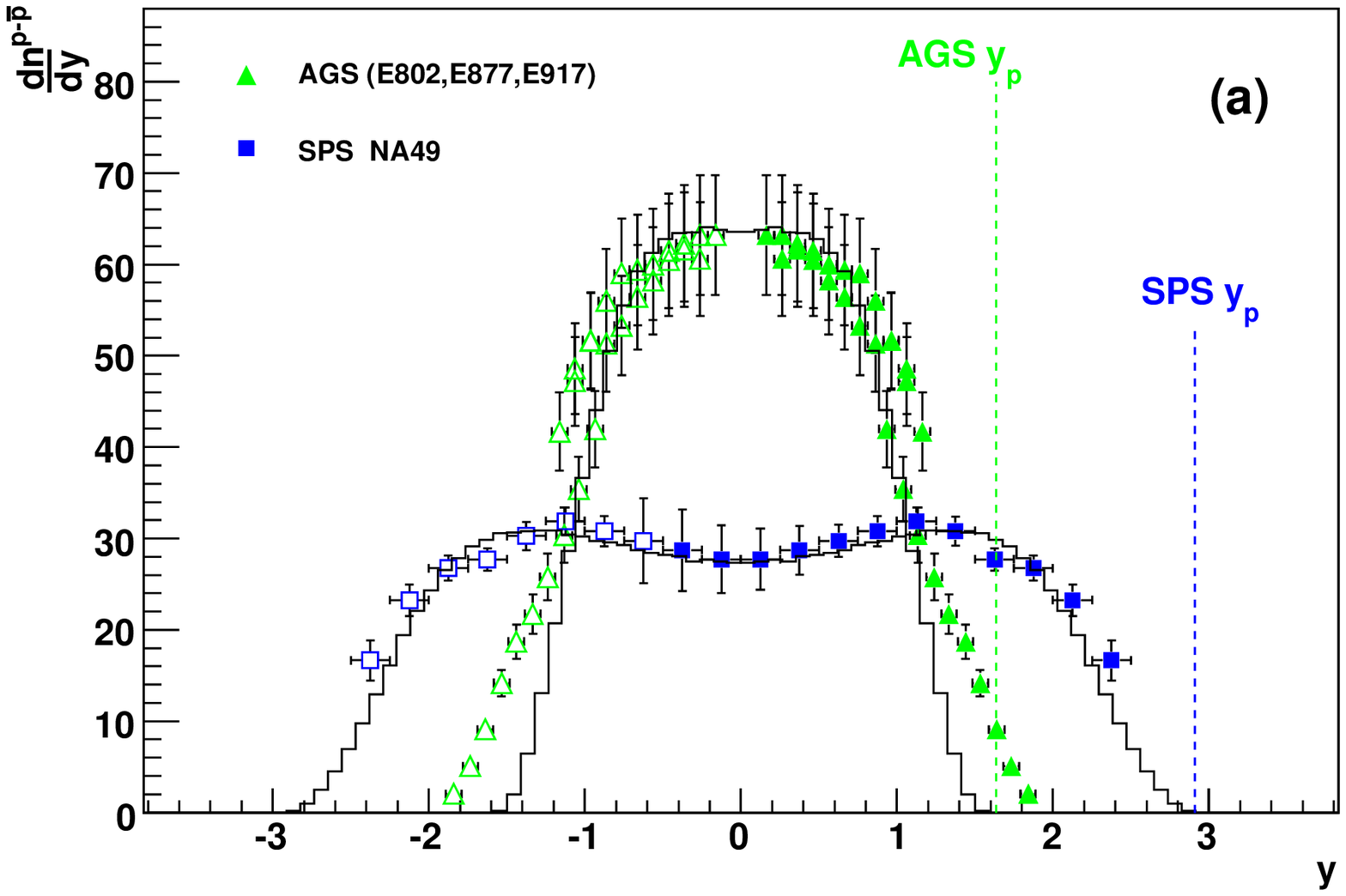, angle=0, width=0.4\textwidth}
\epsfig{figure=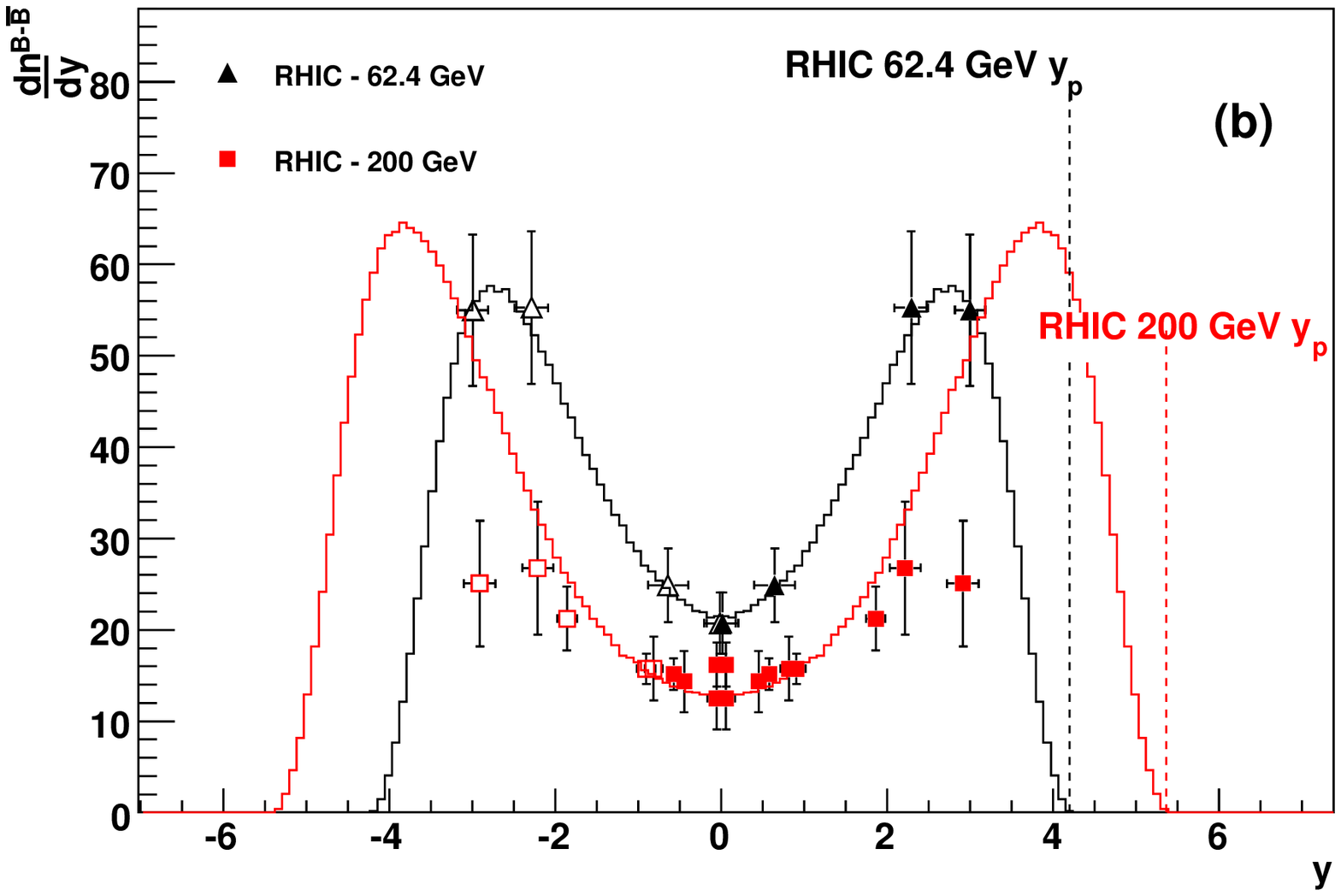, angle=0, width=0.4\textwidth}
\caption{The results of the present model for: (a) net-proton and (b) net-baryon rapidity 
are compared to experimental data from central A-A collisions at different
centre-of-mass energies. See~\cite{rhic-brahms,rhic-brahms2,sps,ags} for details on the data points.}
\label{all-dndy}
\end{center}
\end{figure}

\begin{figure}[htb]
\begin{center}
\fontsize{9pt}{9pt} 
\epsfig{figure=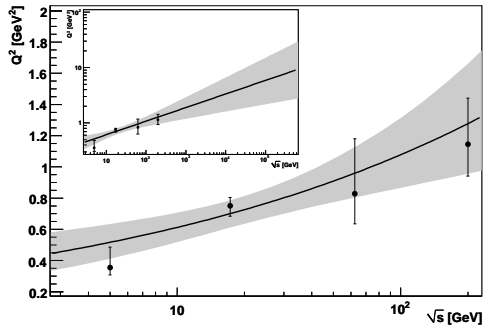, angle=0, width=0.4\textwidth}
\caption{The effective $Q^2_A$ values at different centre-of-mass energies chosen by tuning the model to
the experimental data are shown. The line shows the fit using eq.~(\ref{eq:q-all}) and considering 
Au-Au collisions. The shaded areas correspond to 1$\sigma$ variations of the fit parameters.
The inset figure shows the extrapolation of the fit to higher energies, covering the
LHC and very high energy cosmic rays.}
\label{fit_lambda}
\end{center}
\end{figure}

Turning now to the evolution with $\sqrt{s}$, we consider
the results available for central collisions at RHIC, the SPS
and AGS~\cite{rhic-brahms,rhic-brahms2,sps,sps2,ags}.
As above, in order to fully define the model we need to choose an effective $Q^2_A$ value 
and a normalisation factor related to $N_{part}$ (and to $\varepsilon_s$, in the case of net-proton).
In practice, this was done by fitting the model predictions to the
existing rapidity data using as free parameters the effective $Q^2$ (and assuming the relation 
to $Q^2_A$ we have just inferred from centrality data)
and a normalisation factor $n$.
The data points and the results of the fit are shown in figure~\ref{all-dndy}.
As stated above, the experimental results are presented in terms of net-baryon rapidity in the case of BRAHMS 
and as net-proton rapidity in the case of NA49 and of the AGS experiments. 
The model calculations were performed with equation~(\ref{eq:AA-netB}) for the net-baryon case
and~(\ref{eq:AA-netp}) for the net-proton one.  
A good description of the existing data is achieved.
At  $\sqrt{s} \simeq 5$~GeV only the points up to the nominal beam rapidity were considered,
since our model does not include low energy effects relevant at these energies,
such as Fermi momentum effects, and therefore has no mechanism to reproduce these data. 
At all the other centre-of-mass energies all data points were included in the fit.
The present RHIC data at $\sqrt{s}=200$~GeV cover only the mid-rapidity range, 
leaving the fit largely unconstrained. For this reason, at this centre-of-mass
energy the normalisation factor was fixed to the value estimated in
reference~\cite{rhic-brahms} and only $Q^2$ was left as a free parameter.
In all the other cases the normalisation was left free.
\begin{center}
\begin{table}
\begin{center}
\begin{tabular}{cccccc}
\hline
$\sqrt{s}$ \scriptsize{(GeV)} & Collision & $Q^2$ \scriptsize{(GeV$^2$)} & $n$ & $N_{part}$ & $N_{part}^{Ref}$\\  
\hline
5         & Au-Au   &    $0.35^{+0.13}_{-0.05}$    &  $66.6^{+2.4}_{-3.1}$  & $266.4^{+9.6}_{-12.4}$   & $344 \pm 6$  \\ 
17        & Pb-Pb   &    $0.76^{+0.05}_{-0.07}$    & $67.7^{+1.4}_{-1.4}$  & $361.1^{+7.5}_{-7.5}$    & $362 \pm 12$ \\ 
62.4      & Au-Au   &    $0.77^{+0.33}_{-0.18}$    & $148.2^{+12.6}_{-12.4}$  & $296.4^{+25.2}_{-24.8}$  & $314 \pm 8$  \\ 
200       & Au-Au   &    $1.14^{+0.29}_{-0.20}$    &        -              &       -         & $357 \pm 8$  \\ 
\hline
\end{tabular}
\caption{Results of the fit to the effective $Q^2$ and the normalisation factor $n$ at the different centre-of-mass energies.
The number of participants obtained from equations~(\ref{eq:AA-netB}) and~(\ref{eq:AA-netp}) ($N_{part}$) and estimated
by the experiments ($N_{part}^{Ref}$) are also given.}
\label{tab:results}
\end{center}
\end{table}
\end{center}
The values obtained for the effective $Q^2$, the normalisation factor $n$ and the number of participants 
$N_{part}$ (computed from equations~(\ref{eq:AA-netB}) and~(\ref{eq:AA-netp}))
are given in table~\ref{tab:results}, where the number of participants estimated by the
experiments~\cite{rhic-brahms,rhic-brahms2,sps2,ags} for the relevant centrality region is also given.
The fitted $N_{part}$ values are in good agreement with the expectations.
The discrepancy at $\sqrt{s}=5$~GeV is due to the effect in the normalisation
factor of excluding the points above the nominal centre-of-mass energy from the fit.

As discussed above, strangeness contribution should be simply a global factor when
comparing with net-proton results and was taken to amount to $\varepsilon_s=0.25$
at SPS energies and to be negligible at AGS.
It could slightly distort the spectrum in the comparison with net-baryon rapidity distributions
at RHIC, but this is not yet relevant given the error bars of the available data points.

Assuming equation~(\ref{eq:q-all}) to describe the evolution of the effective $Q^2_A$ with
centrality and $\sqrt{s}$,
the exponent $\lambda_v$ was determined by fitting this equation to the 
$(\sqrt{s},Q^2)$ points in table~\ref{tab:results}. This is shown in figure~\ref{fit_lambda},
where the points are the $Q^2$
values adjusted above, the line is the fit to these points with equation~(\ref{eq:q-all})
and the shaded areas correspond to 1$\sigma$ variations of the fit parameters.
The value obtained for the exponent was $\lambda_v=0.25 ^{+0.12}_{-0.11}$, 
taking $\sqrt{s_0} \simeq 17$~GeV. 

\begin{figure}[htb]
\begin{center}
\fontsize{9pt}{9pt}
\epsfig{figure=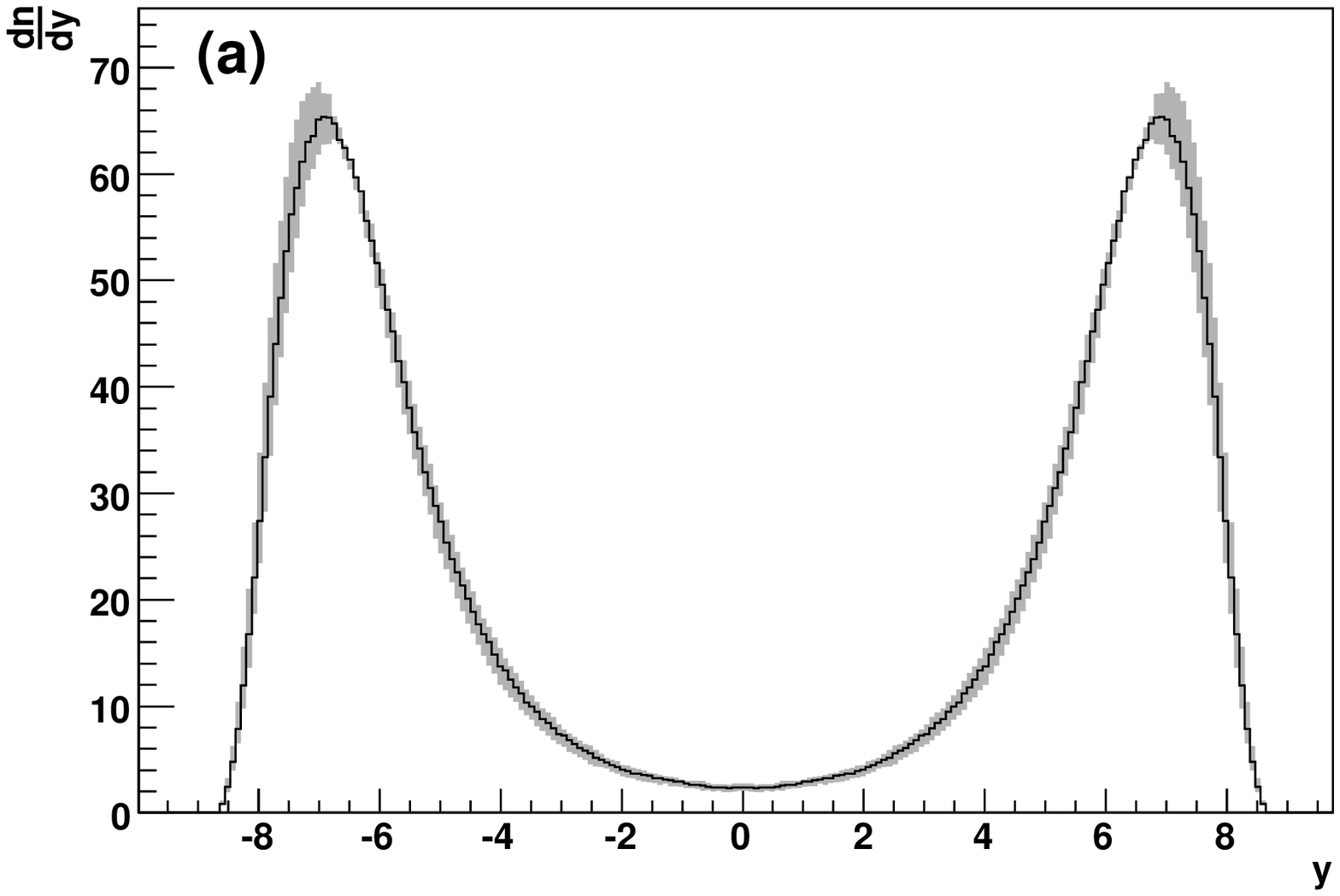, angle=0, width=0.4\textwidth}
\epsfig{figure=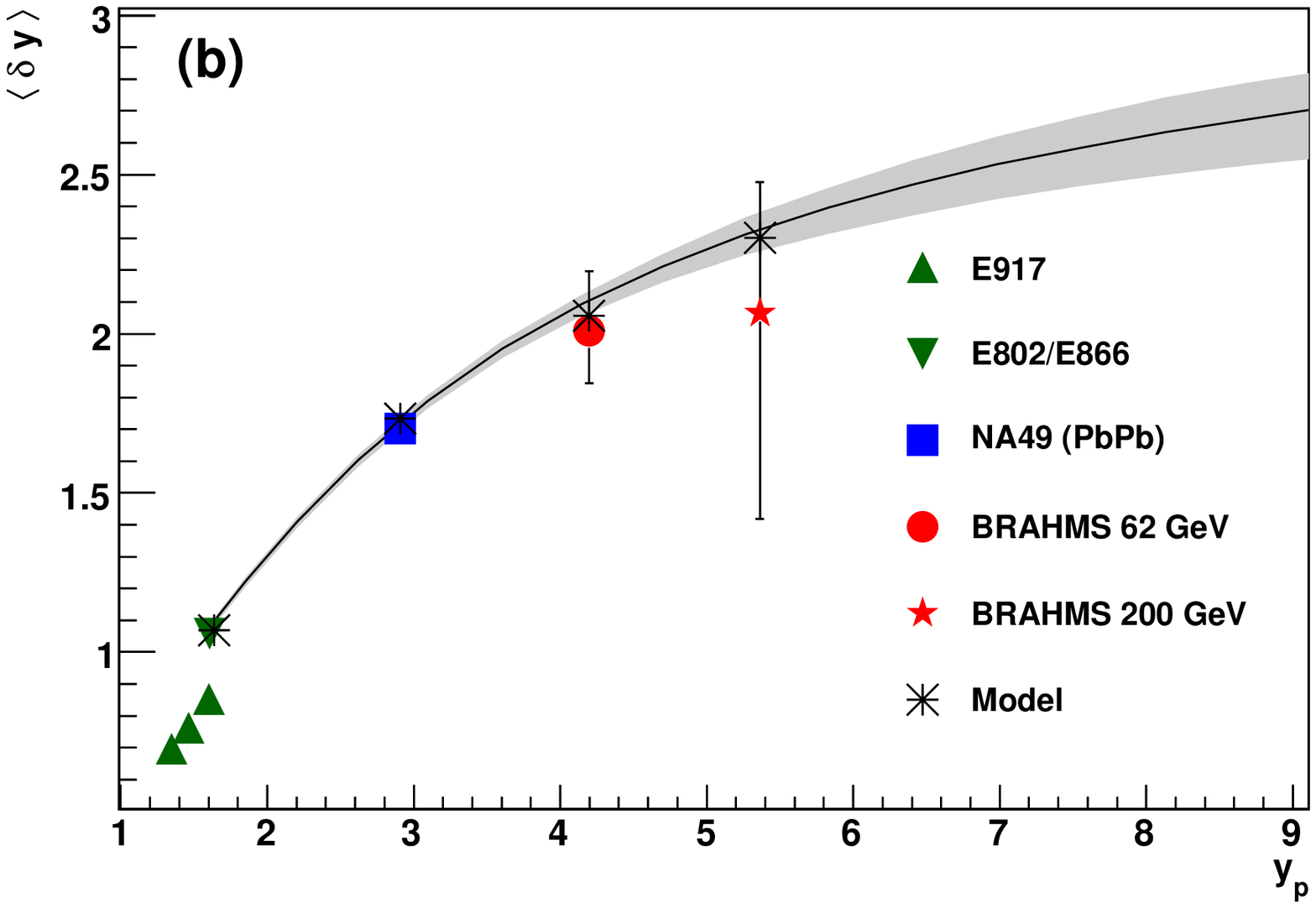, angle=0, width=0.4\textwidth}
\caption{Predictions of the present model for (a) net-baryon rapidity 
for central Pb-Pb collisions at $\sqrt{s}=5.5$~TeV and 
(b) rapidity loss as a function of $\sqrt{s}$. The full line 
is the prediction of the model for Au-Au collisions with a 5\%
centrality cut. The experimental results available at the different
centre-of-mass energies are also shown (see key inside the figure), 
and compared to the predictions of the model for the same type
of collisions and centrality cut (stars).}
\label{predictions}
\end{center}
\end{figure}

\begin{figure}[htb]
\begin{center}
\fontsize{9pt}{9pt}
\epsfig{figure=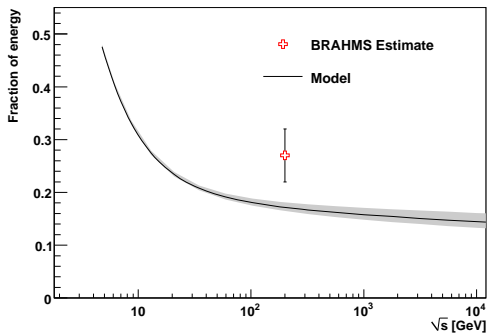, angle=0, width=0.4\textwidth}
\caption{Evolution of the fraction of energy carried by the net-baryon with $\sqrt{s}$.
The prediction of the present model (with a shaded band corresponding to the 1$\sigma$ variation
of the fit parameters) for Au-Au collisions with a 5\% centrality cut are shown and compared 
to the RHIC estimate given in~\cite{Back}.}
\label{lambda-x}
\end{center}
\end{figure}

The predictions of the present model for net-baryon rapidity and rapidity loss 
at higher centre-of-mass energies were then obtained and are shown in
figure~\ref{predictions}.
The width of the curves corresponds to varying the $Q^2$ within 
the 1$\sigma$ band shown in figure~\ref{fit_lambda}.
In figure~\ref{predictions}(a) the net-baryon rapidity distribution
in central Pb-Pb collisions at $\sqrt{s}=5.5$~TeV is presented.  
Following reference~\cite{nardi}, the number of participants 
corresponding to selecting the 6\% most central collisions was set.
In figure~\ref{predictions}(b) the available data points
on the rapidity loss are shown and compared to the predictions of the model.
For the model, two types of results are shown: the full line is 
the prediction for Au-Au collisions with a centrality cut of 5\%,
while the stars show the prediction directly comparable with the
experimental data point at the same $\sqrt{s}$ (same type of collisions
and centrality cut). A good agreement is found in all cases.
The fact that the 1$\sigma$ band is quite narrow in this variable when 
compared to the $Q^2$ one results from the weak dependence on $Q^2$ reached
at high $Q^2$ values (and high energies).

Finally, the fraction of the centre-of-mass energy  
carried by the net-baryon as a function of $\sqrt{s}$ 
was computed and is shown in figure~\ref{lambda-x} (together with the 1$\sigma$ bounds). 
The prediction is for Au-Au central collisions.
According to~\cite{Back}, RHIC data indicate about 27\% of the
initial energy remaining in the net-baryon after the collision. This result is also
shown in the figure.

At RHIC energies, the present model is somewhat below the measured value,
but within 2$\sigma$. 
At higher energies, a sizable amount of 
energy is still associated to the net-baryon.
It should be noted that high energy effects such as string percolation may 
change these predictions~\cite{percol}.

\section{Summary and conclusions}
\label{sec:conclusions}

A simple and consistent model for net-baryon production in high energy 
proton-proton and nucleus-nucleus collisions 
was presented. The basic ingredients of the model are valence string formation based on standard PDFs 
with QCD evolution and string fragmentation via the Schwinger mechanism.
The results of the model were presented and compared with data from RHIC, SPS and AGS.

The free parameters in the model, the effective $Q^2$ and the number of participating
nucleons, were fitted to net-proton data published by AGS experiments and by NA49
and at $\sqrt{s} \simeq 5$~GeV and $\sqrt{s} \simeq 17$~GeV, respectively, and by the
BRAHMS experiment and $\sqrt{s}=62.4$~GeV and $\sqrt{s}=200$~GeV.
A good fit to the data is obtained within this model, with the number of 
participants matching the calculations in the literature. The results show that a good 
description of the main features of net-baryon data is achieved on the basis of this 
simple model, in which the fundamental production mechanisms appear in a transparent way.

Using the available net-baryon data at different centre-of-mass energies and centralities, 
a relation
between the effective momentum scale and the centre-of-mass energy was motivated 
and a prediction was obtained and extrapolated to higher energies for the evolution
with $\sqrt{s}$ and $N_{part}$ of the fraction of the initial energy carried away by the net-baryon.
A sizable amount of energy may be associated to the net-baryon,
even at high energies.

\section*{Acknowledgments}

We thank K. Werner for kindly providing us the EPOS 1.61 code and for 
useful discussions and advice on its usage and results. 
We thank S. Ostapchenko for kindly providing us a QGSJET-II.03
standalone heavy-ion version and for useful discussions.
We thank N. Armesto for useful discussions.
J. A-M thanks Ministerio de Ciencia e
Innovaci\'on (FPA 2007-65114 and Consolider CPAN);
Xunta de Galicia (PGIDIT 06 PXIB 206184 PR) and
Conseller\'\i a de Educaci\'on (Grupos de Referencia Competitivos --
Consolider Xunta de Galicia 2006/51); and Feder Funds, Spain.
R. Concei\c{c}\~ao acknowledges the support of 
FCT, Funda\c{c}\~ao para a Ci\^encia e a Tecnologia and Feder Funds, Portugal.
J. Dias de Deus and J.G. Milhano are partially funded by FCT under 
project CERN/FP/83593/2008.

\end{document}